\documentclass[aps,prB,superscriptaddress,twocolumn,showpacs,amsmath,amssymb]{revtex4-1}

\usepackage{graphicx}
\usepackage{float}

\begin{document}
\title{Photoemission perspective on pseudogap, superconducting fluctuations, and charge order: a review of recent progress}
\author{I. M. Vishik}
\affiliation {University of California, Davis}

\date{\today}

\begin{abstract}
In the course of seeking the microscopic mechanism of superconductivity in cuprate high temperature superconductors, the pseudogap phase\textemdash the very abnormal 'normal' state on the hole-doped side\textemdash has proven to be as big of a quandary as superconductivity itself.  Angle-resolved photoemission spectroscopy (ARPES) is a powerful tool for assessing the momentum-dependent phenomenology of the pseudogap, and recent technological developments have permitted a more detailed understanding.  This report reviews recent progress in understanding the relationship between superconductivity and the pseudogap, the Fermi arc phenomena, and the relationship between charge order and pseudogap from the perspective of ARPES measurements.
\end{abstract}

\maketitle

\section{Introduction}

\begin{figure*}[!]
\includegraphics [type=jpg,ext=.jpg,read=.jpg,clip, width=7 in]{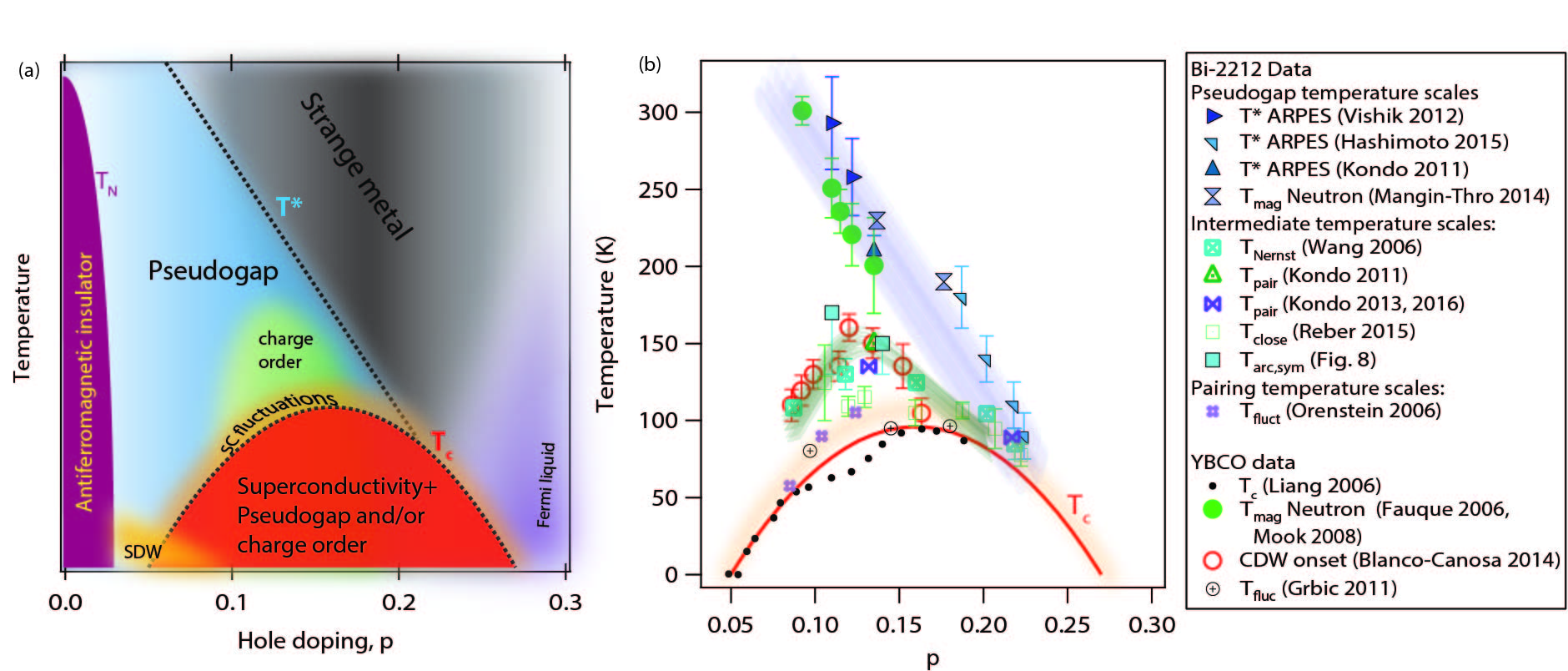}
\centering
\caption[Fig PhaseDiagramSchematic] {\label{Fig PhaseDiagramSchematic} Schematic phase diagram (a) of hole-doped cuprates, based in part on data shown in panel (b) from Refs. \cite{Vishik:PhaseCompetition,Hashimoto:DirectSpectroscopicEvidence,Kondo:DisentanglingCooperPair,Mangin-Thro:CharacterizationIntraUnitCellBi2212,Wang:NernstEffect,Kondo:FormationGaplessFermiArcs,Kondo:PointNodesPersisting,Reber:CoordinationEnergyTemperature,Orenstein:SuperconductingFluctuations,Liang:EvaluationHoleDopingYBCO,Fauque:MagneticOrderPG,Mook:ObservationMagneticOrder,Blanco-Canosa:RIXS_CDW,Grbic:TemperatureRangeSuperconductingFluctuations}. The hole doping is determined from $T_c$ from a universal curve\cite{Presland:UniversalCurve} for Bi-2212 and from a modified version thereof for YBCO\cite{Liang:EvaluationHoleDopingYBCO}.  This review will focus on the pseudogap regime, and its relation to both charge order and superconducting fluctuations.}
\end{figure*}

By virtue of its proximity to the highest temperature superconducting transition found at ambient pressure, the pseudogap in hole-doped cuprates has emerged as the most celebrated `mystery phase' in condensed matter physics.  Almost every experimental probe which couples to electrons observes a change, onset, or crossover at a characteristic temperature $T^*$, which increases with underdoping\cite{Timusk:PseudogapExptSurvey,Tallon:DopingDepT_star,Lee:DopingMottInsulatorReview}.  The theoretical explanations for this phase are equally numerous, including various types of density waves (pair, charge)\cite{Emery:SpinGapProximity_HTSC,Perali:DSC_nearChargeInstability,Lee:AmpereanPairing, Fradkin:IntertwinedOrders}, intra-unit cell orbital currents\cite{Varma:nfl_pairingInstability,Varma:TheoryPG}, and nematic order\cite{Vojta:stripesNematicsSuperconductivity,Fradkin:NematicFermiFluid}.  The intent of this review is not to advocate for a specific explanation for the pseudogap, but rather, to gather recent evidence from photoemission experiments as to the key phenomenology that a theory of the pseudogap must capture.  A schematic phase diagram of hole doped cuprates is shown in Fig. \ref{Fig PhaseDiagramSchematic}, together with the experimental signatures which will be discussed in this review.

\begin{figure}[!]
\includegraphics [type=jpg,ext=.jpg,read=.jpg,clip, width=3.5 in]{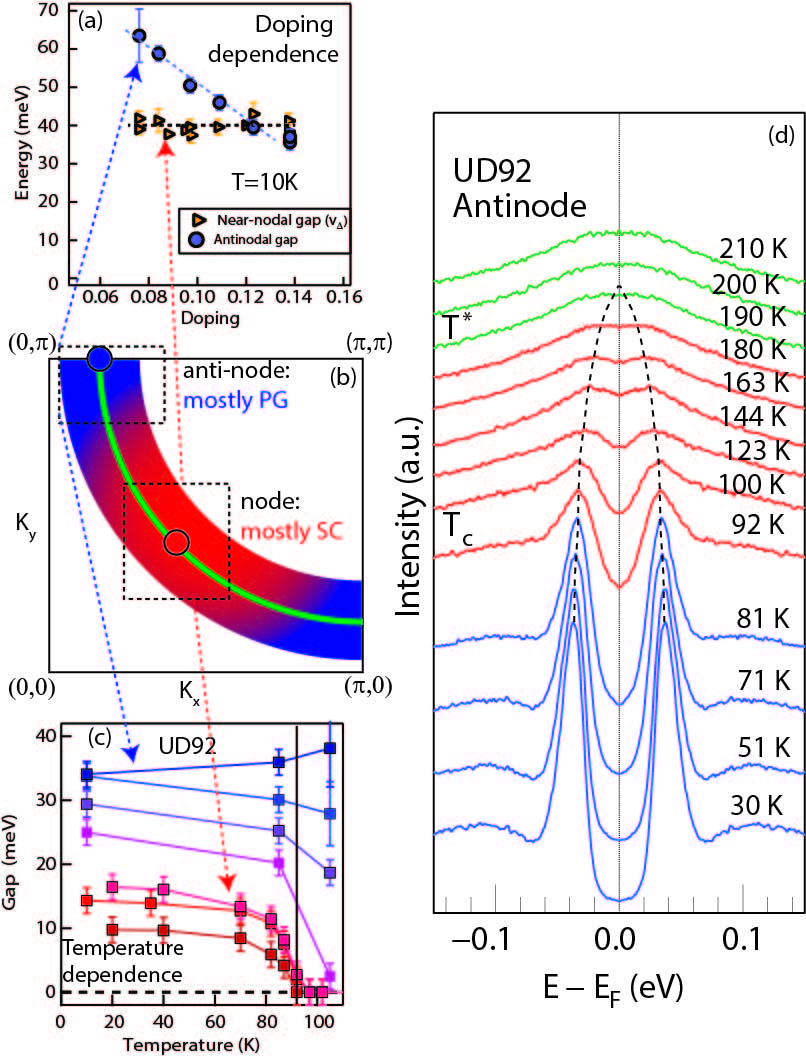}
\centering
\caption[Fig CuprateDichotomy] {\label{Fig CuprateDichotomy} Nodal/antinodal dichotomy with respect to gap measurements (a)Doping dependence of antinodal energy and near-nodal energy scale ($v_{\Delta}$), measured well below $T_c$ and adapted from Ref. \cite{Vishik:PhaseCompetition}. (b) Schematic of nodal-antinodal dichotomy shown on $1/4$ of the BZ with Fermi surface indicated in green.  Distinct phenomenology on different regions of one Fermi surface is interpreted as gaps of two different origins dominating there.  However it should be emphasized that superconducting quasiparticles are observed at the antinode\cite{Vishik:UbiquitousAntinodalQuasiparticles,Vishik:UbiquitousAntinodalQuasiparticles} and the onset of the pseudogap has pronounced signatures in near-nodal phenomenology\cite{Vishik:PhaseCompetition}. (b) Temperature dependence of UD92 ($p\approx0.14$) showing gap closing across $T_c$ near the node and gap persisting across $T_c$ at the antinode.  Adapted from Ref. \cite{Lee:AbruptOnset} (c) Temperature dependent symmetrized energy distribution curves (EDCs) at the antinode.  $T^*$ is commonly identified in ARPES data as the temperature where symmetrized spectra merge into one peak.  From Ref. \cite{Hashimoto:EnergyGaps}}
\end{figure}

ARPES is a crucial technique for learning about cuprate high temperature superconductors because the key emergent phases in these materials--\textit{d}-wave superconductivity and the pseudogap--are characterized by single-particle gaps whose magnitudes are anisotropic in momentum space, reaching their maximum magnitude at the Brillouin zone (BZ) boundary.  Specifically, the pseudogap is marked in ARPES experiments by a gap which develops at high temperature.  The first appearance of this gap at the antinodal momentum, located at the boundary of the BZ, is a common definition of T$^*$ from ARPES (Fig. \ref{Fig CuprateDichotomy}(d)).  Since the early 2000s, technological improvements in the technique have facilitated the acquisition of more detailed data which challenges prior understanding of the pseudogap.  Important for the present discussion, improved energy resolution, now routinely better than 10 meV, has permitted the investigation of smaller gaps in the near-nodal regions to better accuracy.  Less appreciated is the increasing brightness of photoemission lightsources--laser, synchrotron, and gas discharge lamp--which permits collecting data with better statistics, measuring more temperature and momentum data points, and finishing  the measurement faster before sample aging ensues.  The consequences of these technological improvements are apparent in the recent insights emerging from many research groups.  This review focuses on ARPES results on $Bi_2Sr_2CaCu_2O_{8+\delta}$ (Bi2212), a bi-layer cuprate whose pristine cleaved surface and sharp lineshapes, and high $T_c$ makes it an ideal material for surface-sensitive spectroscopies.  While other cuprates can be synthesized to have much longer electron mean free paths, those materials typically yield broader ARPES spectra, making them poorer candidates for some of the quantitative lineshape analysis used to assess pseudogap physics.

Reviews of recent gap measurements via ARPES are presented elsewhere\cite{Vishik:ARPESCuprateFermiology,Hashimoto:EnergyGaps,Kordyuk:PseudogapFromARPES}, and serve as a starting point for this review. A distinct phenomenology between the near-nodal and the near-antinodal region of the Brillouin zone is observed, and this is interpreted in terms of a 'two gap' picture, where the pseudogap is a distinct electronic phase from superconductivity, with the two coexisting below T$_c$.  In the superconducting state, the pseudogap dominates near the antinode, while superconductivity dominates near the node, though superconducting features are present all around the Fermi surface\cite{Vishik:UbiquitousAntinodalQuasiparticles}.  This dichotomy, as it relates to the doping and temperature dependence of spectral gaps, is sketched in Fig. \ref{Fig CuprateDichotomy}.  The distinct nature of the pseudogap and superconductivity has myriad support from ARPES data including the following: distinct doping, temperature, and momentum dependence of gaps on a single Fermi surface\cite{Lee:AbruptOnset,Vishik:PhaseCompetition}, particle-hole symmetry breaking in the antinodal region below $T^*$\cite{Hashimoto:ParticleHoleSymmetryBreaking}, and a non-monotonic evolution of spectral weight as a function of momentum and temperature\cite{Kondo:CompetitionPGSC,Hashimoto:DirectSpectroscopicEvidence}. The 'two gap' picture is corroborated by a number of other experiments which suggest a symmetry breaking transition distinct from superconductivity at $T^*$.  Some examples include the appearance of intra-unit cell magnetic order in neutron scattering\cite{Fauque:MagneticOrderPG,Mook:ObservationMagneticOrder} (labeled $T_{mag}$ in the phase diagram in Fig. \ref{Fig PhaseDiagramSchematic}), a break in slope of the temperature dependence of elastic moduli measured by resonant ultrasound experiments\cite{Shekhter:BoundingPseudogap}, and broken inversion and rotation symmetry as reveal by nonlinear optics\cite{Zhao:globalInversionSymmetryBroken}.

More recently, another character has emerged prominently on the phase diagram--charge density wave (CDW) order or short-range correlations thereof.  Although a CDW, particularly one that is short range, should share spectral features with the pseudogap, it appears that the CDW in hole doped cuprates has distinct phenomenology from the pseudogap: it has lower onset temperature and a distinct doping dependence.

This report is structured in the following way.  First, it addresses differing pseudogap phase diagrams based on different experimental techniques on different materials, and then it discusses some of the subtle spectral features that need to be accounted for in explaining the pseudogap.  Afterwards, it moves to the near-nodal region to discuss Fermi arcs and new fitting methodologies which point to their absence above $T_c$.  The last sections touch on some of the features in ARPES data which may be related to the CDW, and others that are probably not.

\section{The end of the pseudogap}
There is substantial evidence of a zero-temperature phase boundary inside the superconducting dome on the slightly overdoped side at $p\approx18-20\%$ hole doping.  Among the experiments showing support for this are quantum oscillation measurements which show a diverging effective mass approaching this doping from below\cite{Ramshaw:QPMassEnhancement}, scanning tunneling spectroscopy (STS) quasiparticle interference (QPI) implying the sudden appearance of a 'large' Fermi surface\cite{Fujita:SimultaneousTransitionsCuprate}, a maximum in the superfluid density\cite{Tallon:SuperfluidDensity}, and high-field thermal transport experiments showing both a local maximum in $H_{c2}$\cite{Grissonnanche:DirectMeasurementHc2} and a change in carrier density\cite{Boebinger:InsulatorMetalCrossoverLSCO,Laliberte:OriginMetaltoInsulator} (with the crossover doping range being somewhat material-dependent).  In ARPES experiments, evidence for this phase boundary at low temperature come from a sudden change in the phenomenology of the near-nodal gap in the superconducting state from doping-independent to doping-dependent\cite{Vishik:PhaseCompetition}, and earlier, from the doping dependence of the antinodal quasiparticle weight\cite{Feng:SuperfluidDensityARPES}.  Whether the observed low-temperature phase boundary in various experiments originates from  charge order or from the pseduogap is still a matter of discussion.  It should be noted that not all experiments support a phase boundary inside the superconducting dome on the overdoped side, in particular, recent measurements of superfluid density in ultra-clean $La_{2-x}Sr_xCuO_4$ (LSCO) films do not reproduce earlier anomalies at 19$\%$ hole doping \cite{Bozovic:SuperfluidDensityODLSCO}.

Above $T_c$, the endpoint of the pseudogap is more disputed, with one version of the phase diagram having $T^*$ plunging into the superconducting dome around optimal doping and another version having it intersect the superconducting dome on the overdoped side\cite{Tallon:DopingDepT_star}.  This is a complex question to unravel because these two versions of the phase diagram have support from different compounds which are synthesizeable in different doping ranges and conducive to different experimental techniques. For this reason, the phase diagram in Fig. \ref{Fig PhaseDiagramSchematic}(b) separates data collected on $YBa_2Cu_3O_{6+x}$ (YBCO) and Bi2212, two cuprates whose data are frequently compared because they are both bi-layer systems with similar $T_c$.

The former phase diagram, in which a pseudogap is absent above $T_c$ on the overdoped side, is supported by data on YBCO, which is a compound favored for scattering and transport experiments, but difficult for ARPES because of surface self-doping and spectral contamination from chain layers\cite{Hossain:InSituDopingControl}. In YBCO, high temperature measurements of $T^*$ appear to extrapolate to a $p\approx0.19$ endpoint supported by low temperature experiments, including data points below $T_c$\cite{Shekhter:BoundingPseudogap}.

The latter picture of the phase diagram, in which the pseudogap is present on the overdoped side at least to $p\approx0.22$, is often dubbed 'the spectroscopist's phase diagram' because ARPES, STS, and Raman scattering all show a gap in the normal state in the overdoped regime in Bi2212, qualitatively similar to the gap observed on the underdoped side\cite{Gomes:VisualizingPairFormation, Vishik:PhaseCompetition, Benhabib:CollapseNormalStatePseudogap}. In ARPES, $T^*$ is commonly determined via symmetrization—mirroring energy distribution curve (EDC) at $k_F$ about $E_F$ and adding it back to itself \cite{Norman:DestructionFermiSurface}.  If there is particle-hole symmetry at $k_F$, this procedure removes the Fermi-Dirac cutoff, but in the absence of or ignorance about particle-hole symmetry it serves as a visualization tool for the gap on the occupied side: finite gap gives two peaks and zero gap gives one.  In the absence of particle-hole symmetry in the pseudogap regime, symmetrization can underestimate $T^*$ if the gap is centered above $E_F$.  Other assessments can be used to quantify $T^*$ including leading edge midpoint, integrated momentum distribution curve (MDC) weight at $E_F$ \cite{Kondo:FormationGaplessFermiArcs}, dividing EDCs at $k_F$ by a resolution-convolved Fermi-Dirac function \cite{Hashimoto:DirectSpectroscopicEvidence}, and temperature-dependent integrated EDC spectral weight\cite{Hashimoto:DirectSpectroscopicEvidence}. In general, these procedures yield values of $T^*$ within 30K of one another when performed on the same dataset, with symmetrization giving a lower value.

A number of reconciliations have been offered for the differing phase diagrams with respect to the $T^*$ lines.  One possibility is a speculated re-entrant pseudogap\cite{Vishik:PhaseCompetition}.  Neutron scattering experiments in Bi2212 may provide an alternate explanation. These studies identify the pseudogap via intra-unit cell magnetic order, and also show that the characteristic intra unit cell magnetic order becomes increasingly short range in the overdoped regime\cite{Mangin-Thro:CharacterizationIntraUnitCellBi2212}.  This can be interpreted as remnant underdoped or optimally-doped patches inside a heterogenous sample contributing to the spectroscopic pseudogap in the overdoped regime, which may also explain the seeming kink in $T^*$ determined by ARPES around optimal doping.  Another proposal is that the normal-state spectral gap above $T_c$ is different on the underdoped and overdoped side, with the latter being attributed to superconducting fluctuations\cite{Gomes:VisualizingPairFormation,Zaki:CupratePhaseDiagram}. However, this interpretation contrasts with other ARPES experiments in overdoped Bi2212 that observe \textit{the same} signature of the pseudogap in underdoped and overdoped samples.  For example, Ref. \cite{Hashimoto:DirectSpectroscopicEvidence} reports non-monotonic temperature dependence of spectral weight that is consistent with phase competition between superconductivity and the pseudogap on either side of optimal doping\cite{Hashimoto:DirectSpectroscopicEvidence}.  A pseudogap of identical origin and properties in slightly underdoped and slightly overdoped Bi2212 would suggest that the different T* phase diagrams may simply originate from unidentified materials differences between YBCO and Bi2212 which lead to a more robust pseudogap in overdoped Bi2212.

Chemistry presents an additional complexity in ascertaining and interpreting the cuprate phase diagram. Many studies in overdoped Bi2212 make use of Pb-doped samples which tend to provide more stable doping in that regime.  Pb-doping on the Bi site is known to alleviate the buckling of the BiO planes, which greatly diminishes the Umklapp bands and make certain types of data interpretation more straightforward.  However, the role of this chemistry change in affecting the pseudogap has not been quantified, and may further aid in clarifying the fate of the pseudogap on the overdoped side above $T_c$.  YBCO is difficult to dope deep into the overdoped regime, because full chain-oxygenation corresponds to $p=0.194$\cite{Liang:EvaluationHoleDopingYBCO}, so comparison between these two canonical bilayer cuprates ceases beyond the purported low temperature endpoint of the pseudogap.  In the deeply underdoped regime, chemical stability is typically achieved with cation doping on the Ca-site, a procedure that is not without controversy\cite{Zhao:UniversalFeaturesPhotoemission}.  Nevertheless, in doping regimes with overlapping chemistry, ARPES spectra appear to be consistent, and there is a smooth evolution of behavior as one enters the doping regime where cation-doped specimens are favored \cite{Vishik:UbiquitousAntinodalQuasiparticles}.  More broadly, distinguishing universal from materials-dependent properties is a long-standing challenge in cuprates, compounded by the hurdle that the doping in many cuprates, including Bi2212, is not precisely known, but surmised from T$_c$ using an empirical `universal curve' \cite{Presland:UniversalCurve}.  A detailed discussion of this issue is outside the scope of this review, but phase diagrams for other cuprates and comparisons between families can be found elsewhere \cite{Tallon:DopingDepT_star, Ando:ElectronicPhaseDiagram, Li:UnusualMagneticOrder,Daou:TemperatureDependenceResistivity,Hufner:TwoGapsHTSC,Kondo:DisentanglingCooperPair}.

As for the underdoped edge of the pseudogap, it appears that it does not smoothly fade into the antiferromagnetic parent compound.  Neutron scattering measurements on YBCO have reported a sharp decrease in $T_{mag}$ below a doping of $p\approx0.085$, which coincides in doping to the onset of slowly fluctuating short-range spin density wave (SDW) correlations\cite{Baledent:EvidenceCompetingMagneticInstabilities}.  In ARPES experiments, a low-temperature phase boundary is observed at a similar doping, marked by the opening of a gap at the nodal momentum which is present below and above $T_c$\cite{Vishik:PhaseCompetition,Shen:FullyGapped,Razzoli:EvolutionNodelessGap}.  In LSCO, this gap has been directly linked to incommensurate SDW order\cite{Drachuck:SpinChargeInterplayLSCO}.  ARPES measurements have not confirmed whether this phase affects the antinodal pseudogap, in part $T^*$ is very high in this doping range (doping in the top few unit cells tends to be unstable at high temperature\cite{Palczewski:ControllingCarrierConcentration}) and in part because the spectra in Bi2212 become increasingly broad such that it is hard to identify features and their energy scales\cite{Vishik:PhaseCompetition}.

\section{More than a gap}
\begin{figure*}[!]
\includegraphics [type=jpg,ext=.jpg,read=.jpg,clip, width=7 in]{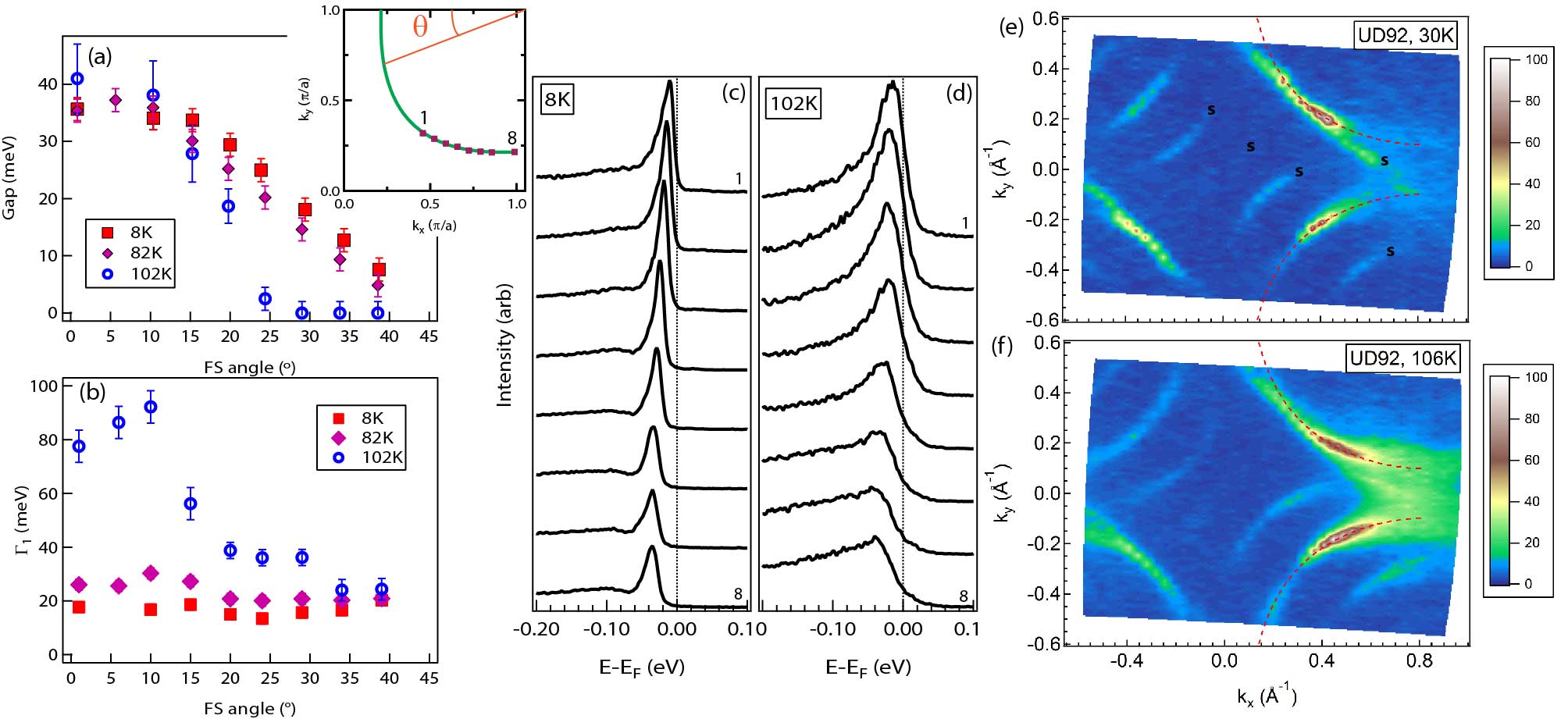}
\centering
\caption[Fig 2] {\label{Fig 2} Subtleties in analyzing pseudogap from ARPES data.  All data on UD Bi-2212 with $T_c=92K$. (a) Gap as a function of Fermi surface angle, defined in inset, below and above T$_c$, adapted from Ref. \cite{Lee:AbruptOnset}. (b) Scattering rate ($\Gamma_1$ from Norman model) at the same three temperatures, showing larger jump across $T_c$ approaching the antinode ($\theta=0$). (c)-(d) EDCs at $k_F$ at the points around FS marked in inset of (a) at 8K (below $T_c$) and at 102K (above $T_c$). (e)-(f) Fermi surface map at 30K and 106K.  Mapping integrates $\pm5 meV$ around $E_F$.  Dashed lines are guide-to-the-eye for Fermi surface (bonding band), and 's' denotes superstructure diffraction replicas, originating from slight corrugation in BiO planes\cite{Fretwell:FermiSurfaceBi2212}.}
\end{figure*}

The pseudogap as measured by ARPES is commonly characterized by a single number--the magnitude of the gap at every momentum around the Fermi surface.  This metric, summarized in Fig. \ref{Fig 2}(a), reveals a gap that persists in the antinodal region above $T_c$, changed little from its value in the superconducting state, and a gapless arc centered around where the node used to be.  This simplification, while convenient in many cases, obscures some of the complexities which may be relevant in connecting experiments to theory. This section aims at highlighting three subtleties that are often hidden when data are described only by the gap: lineshapes as a function of doping, lineshapes across T$_c$, and finite density of states at the Fermi level.  This section will focus on the antinodal region, and later sections will discuss the near-nodal region.

\begin{figure}[!]
\includegraphics [type=jpg,ext=.jpg,read=.jpg,clip, width=3 in]{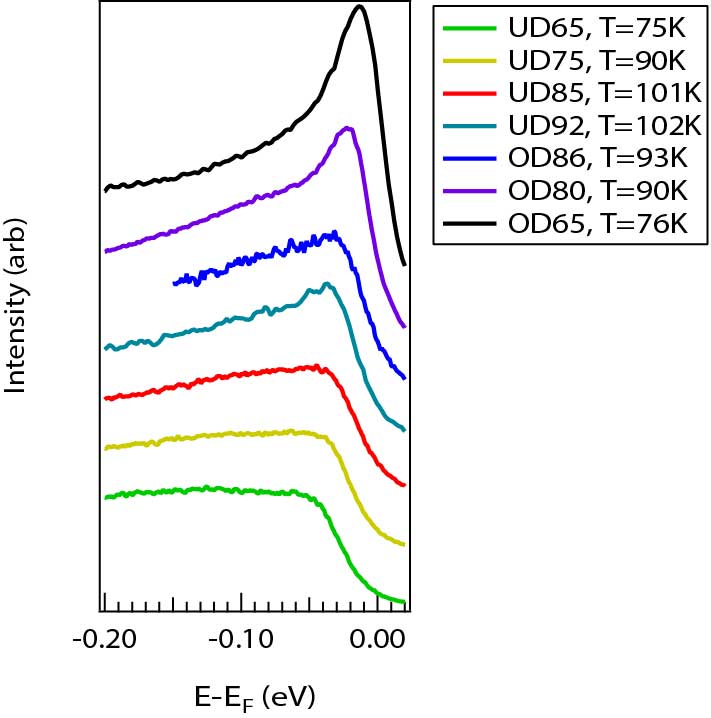}
\centering
\caption[Fig 1] {\label{Fig 1} EDCs taken at antinodal $k_F$ ~10K above $T_c$ for different dopings of Bi-2212.}
\end{figure}

Although the pseudogap temperature decreases with hole doping, it becomes increasingly easier to assign an energy scale to the normal-state gap, because spectra get progressively sharper, as shown in Fig. \ref{Fig 1}.  This is a bit perplexing, given that lower dopings have higher $T^*$, and higher onset temperatures usually imply more robust order.  This expectation is borne out in other characterizations of the pseudogap.  For example, neutron scattering measurements report a larger spin-flip signal in the pseudogap phase in more underdoped samples\cite{Fauque:MagneticOrderPG,Li:UnusualMagneticOrder}.  The observed doping evolution of normal-state ARPES lineshapes was previously discussed in terms of quasiparticle coherence onsetting on the overdoped side\cite{Chatterjee:ElectronicPhaseDiagram}.  Other possibilities include the normal state gap on the overdoped side having a different origin (e.g. superconducting fluctuations vs a distinct ordered state), or underdoped spectra reflecting a greater degree of inhomogeneity broadening\cite{Zaki:CupratePhaseDiagram}.  The latter is supported by STS experiments showing a broader distribution of local gaps in the underdoped regime\cite{Alldredge:EvolutionElectronicExcitation}.

The lineshape in cuprates across T$_c$ is another important factor to consider, specifically the momentum dependence thereof.  Much has been written about the disappearance of a 'peak-dip-hump' lineshape at the antinode across T$_c$ in bilayer cuprates, as well as the general sudden broadening of spectra, and many example spectra are shown in Ref. \cite{Hashimoto:DirectSpectroscopicEvidence}.  The spectral change across $T_c$ is more pronounced in the antinodal region, as shown in Fig. \ref{Fig 2}(b).  For a slightly underdoped Bi2212 sample (UD92), spectra broaden by a factor of 1.5 across $T_c$ in the nodal region, but a factor of 3 in the antinodal region.  A further discussion of quantifying ARPES lineshapes is given in the next section.

Finally, the pseudogap regime is marked by a finite density of states inside the gap, as discussed in detail elsewhere\cite{Kondo:DisentanglingCooperPair}.  This is seen in the constant energy maps in figure \ref{Fig 2}(e)-(f), where (e) is in the superconducting state and (f) is taken above $T_c$.  The latter is marked by substantial intensity in the antinodal region, such that one could not conclude there was a gap there without looking at EDCs (\ref{Fig 2}(d)). One consequence of this is that it is still conceivable to have Fermi surface instability such as superconductivity or charge order originating from the antinode below $T^*$, as there are electronic states at $E_F$ in the pseudogap regime. Another consequence is that the antinodal region can contribute to normal-state transport as well. In any case, it is quite perplexing that the pseudogap has sharp signatures in certain experiments, but is yet marked by broad spectra and finite density of states, and this is something that potential explanations for the pseudogap, or materials' dependence thereof, must capture.

\section{Fitting ARPES data}
There are multiple ways of quantifying ARPES spectra, but perhaps the most widely used is a minimal model proposed by M. Norman \cite{Norman:NormanModel} (Norman Model).

The Norman model is given by the following expression for self energy:
\begin{equation}
\label{Norman model}
\Sigma(\mathbf{k},\omega)=-\textit{i}\Gamma_1+\frac{\Delta^2}{\omega+\epsilon(\mathbf{k})+\textit{i}\Gamma_0}
\end{equation}

$\Gamma$$_1$ is said to be an approximation of the single-particle scattering rate, as the real scattering rate likely depends on energy. $\Gamma$$_0$ is an elastic term that was originally only introduced above T$_c$, associated with an inverse pair lifetime.  Note that this expression is only valid at k$_F$.  Although this model was initially envisioned for a superconductor or a disordered superconductor, it provides a good description of ARPES data on cuprates even in the absence of superconductivity because it has few free parameters.  When applied agnostically to symmetrized spectra at $k_F$, the $\Delta$ term describes the distance from each peak to $E_F$, $\Gamma_1$ describes the width of the peaks, and $\Gamma_0$ describes the nonzero density of states at $E_F$.

\begin{figure}[!]
\includegraphics [type=jpg,ext=.jpg,read=.jpg,clip, width=3.0 in]{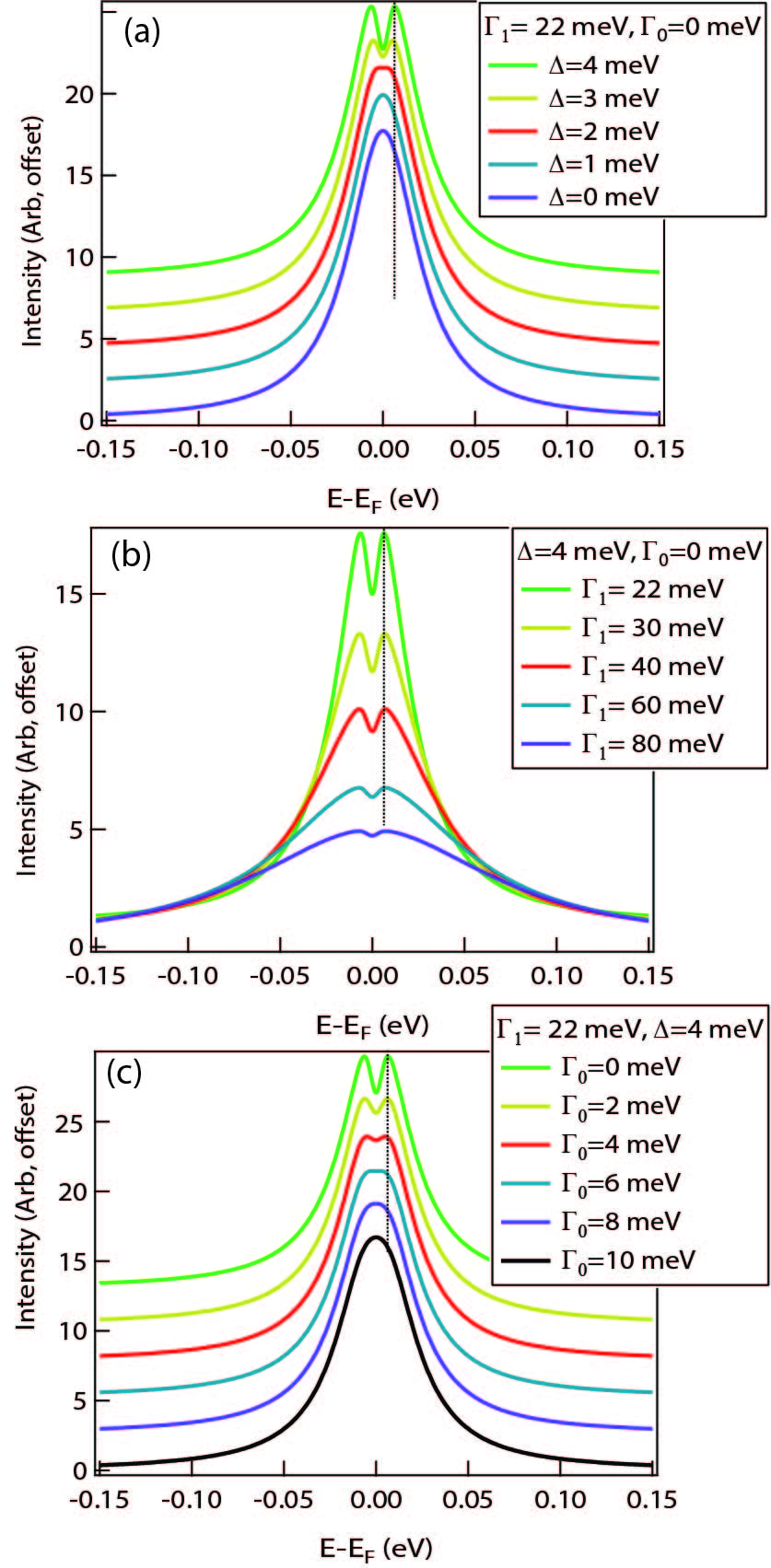}
\centering
\caption[Fig 0] {\label{Fig 0} Simulated single particle spectral function at $k_F$ from Norman model, convolved with an energy resolution of 3 meV. (a) varying $\Delta$ (b) varying $\Gamma_1$ (c) varying $\Gamma_0$.  Vertical line in all panels is guide-to-the-eye for peak position of top trace (green)}
\end{figure}

Fig. \ref{Fig 0} illustrates how each of the parameters in Eqn. \ref{Norman model}, together with instrument energy resolution (taken to be 3 meV), affects the symmetrized EDC at k$_F$.  For simplicity, the simulated EDCs assume particle-hole symmetry which may not be valid in the pseudogap state.  The starting point for each of the panels ($\Delta$$=$4 meV, $\Gamma$$_1$$=$22 meV, $\Gamma$$_0$=0 meV) was chosen to be representative of ARPES data on Bi2212 near the node in the superconducting state.  When $\Delta$ decreases (Fig. \ref{Fig 0}(a)), the peaks on either side of E$_F$ move towards each other, eventually forming a single peak when $\Delta$$=$0.  Increasing $\Gamma$$_1$ (Fig. \ref{Fig 0}(b)) widens the outer envelope of the EDC, but a dip remains at E$_F$ even for $\Gamma$$_1$ which is quite large. In an experiment, a dip at E$_F$ is interpreted as a finite gap. Increasing $\Gamma$$_0$   (Fig. \ref{Fig 0}(c)) fills in spectral weight at $E_F$ without widening the outer envelope.  Notably, when $\Gamma$$_0$ is sufficiently large, there is a single peak in the symmetrized EDC, even though a finite gap was explicitly included in the simulation.  The condition for this is is given in Ref. \cite{Norman:NormanModel} to be 2$\Delta$$^2$$/$$\Gamma$$_0$$^2$+$\Gamma$$_1$$\Delta$$^2$$/$$\Gamma$$_0$$^3$$=$1. In an experiment, such a spectrum would be interpreted as having zero gap.

\section{The vanishing Fermi arcs}
In momentum space, the pseudogap state above T$_c$ is visualized as four disconnected `Fermi arcs' centered around the momenta where the nodes of the \textit{d}-wave superconducting gap existed below T$_c$, accompanied by persistent gaps near the momenta where the antinodes of the superconducting gap once existed\cite{Norman:DestructionFermiSurface}.  The Fermi arcs, if they are to literally be believed as disconnected segments of gapless excitations, are highly anomalous, because a Fermi surface is supposed to be a closed contour.

The Fermi arcs have been attributed to a number of physical entities in the literature.  One proposal is that each Fermi arc constitutes a portion of a closed hole pocket with weak coherence factors on the back side of the pocket\cite{Chakravarty:ARPESdDensityWave,Stanescu:FermiArcs,Yang:EmergencePreformed}, such that only one side is seen in ARPES experiments.  Though there have been some experimental reports of 'oval-shaped' hole pockets expected in this scenario, these results, particularly in Bi2201, appear to be attributable to structural effects--a lesser-known supermodulation of the BiO planes\cite{King:StructuralOrigin,Rosen:SurfaceEnhancedChargeDensityWave}.  In contrast, quantum oscillation experiments in YBCO have given support for a small diamond-shaped \textit{electron}-pocket in the underdoped regime\cite{Sebastian:TowardsResolutionFermiSurface}, originating from Fermi surface reconstruction via charge order which becomes long-range and three dimensional in a magnetic field\cite{Gerber:3DChargeDensityWave}.  Another physical interpretation of the Fermi arcs is that it constitutes the \textit{only} portion of a putative large fermi surface that supports superconductivity\cite{Oda:HyperbolicDependence}.

While the phenomenology of a Fermi arc and an antinodal gap is quite unusual, one well-known ordered state that it does vaguely resemble is a charge density wave (CDW).  With imperfect nesting, it is perfectly reasonable to have part of the Fermi surface be gapped and another part be ungapped.  Though a CDW with imperfect nesting is expected to produce pockets, not arcs, spectra which resemble disconnected arcs have been observed in conventional CDW systems\cite{Moore:FermiSurfaceEvolution}, because of coherence factors on the other side of the pocket.  Another suggested connection between Fermi arcs and charge order is that this charge order is driven by nesting of the \textit{tips} of the Fermi arcs\cite{Comin:ChargeOrderFermiArc}.

However, many of these proposals have difficulty describing the observation that the apparent length of the Fermi arc grows with increasing temperature\cite{Kanigel:EvolutionPseudogap}.  If the length of the Fermi arc is taken to be a physical, doping-dependent quantity, its length should not be strongly temperature dependent, except possibly very close to a phase transition.  Also arguing against the 'small-pocket' class of proposals, ARPES experiments give strong support for a large fermi surface underlying superconductivity because quasiparticles are observed in the superconducting state all around a putative large hole pocket\cite{Vishik:MomentumDependentPerspective}.  Not only does the Fermi arc grow with temperature, but it has been shown to appear even in the superconducting state in impurity-doped Bi2212\cite{Sato:FermiArcImpurity}.  Additionally, ARPES experiments with better instrument resolution tend to show shorter Fermi arcs for a given doping and temperature.

\begin{figure}[!]
\includegraphics [type=jpg,ext=.jpg,read=.jpg,clip, width=2 in]{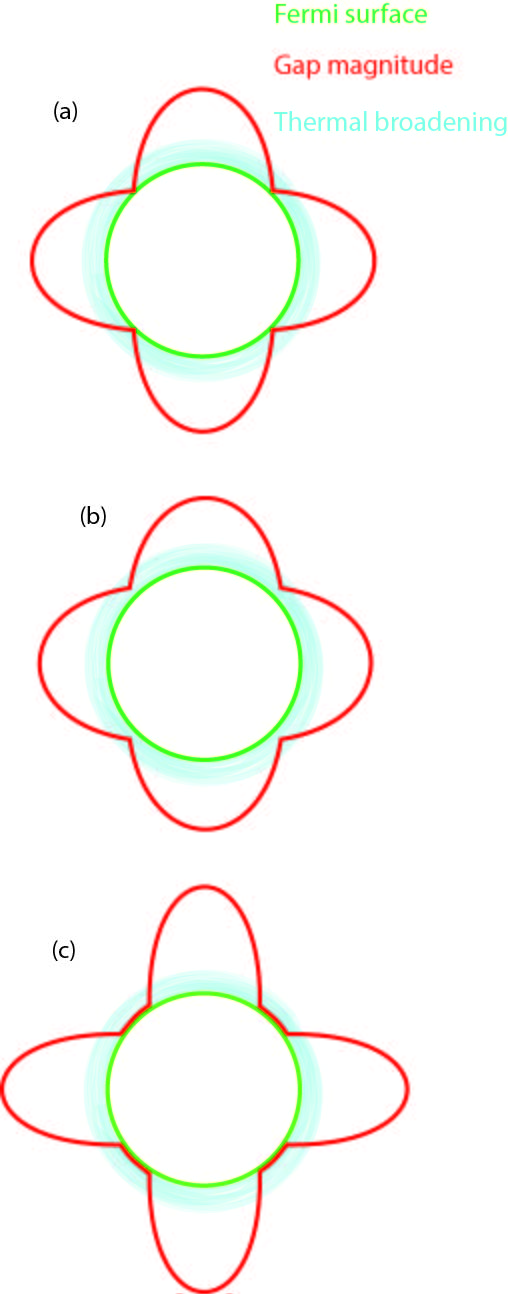}
\centering
\caption[Fig scenarios] {\label{Fig scenarios} Sketch of potential scenarios for normal-state gap structure that would be difficult to distinguish based on symmetrization analysis because of thermal filling of small-energy gaps. (a) point nodes (b) small gap at nodal momentum, as has been observed in deeply underdoped cuprates \cite{Vishik:PhaseCompetition} (c) real arcs}
\end{figure}

Fig. \ref{Fig 0} gets to the heart of the difficulty of assessing the veracity of Fermi arcs.  Although the same gap is always input into the model, an increasing elastic scattering term ($\Gamma_0$) fills in the spectral weight at $E_F$, eventually yielding a single peak in the symmetrized EDC. This is the reason why the temperature dependence of the Fermi arc length may be artificial. In experiments, temperature can play this same role of filling in and obscuring the gap. Bi-2212 has a superconducting transition temperature of almost 100K at optimal doping, lending almost 9 meV of thermal broadening to any ARPES measurement in the normal state.  Convolving instrument resolution  with this spectral function moves the EDC peaks towards $E_F$, as if the gap were closing.  Using symmetrization and inspection alone, one cannot distinguish real Fermi arcs, a point node, and a small gap at the nodal point, as temperature smearing would produce an apparent arc in all of these scenarios (Fig. \ref{Fig scenarios}).  However, new analysis techniques in recent years have permitted progress on this front, and they appear to show a point node persisting until a higher temperature.

\begin{figure}[!]
\includegraphics [type=jpg,ext=.jpg,read=.jpg,clip, width=3.5 in]{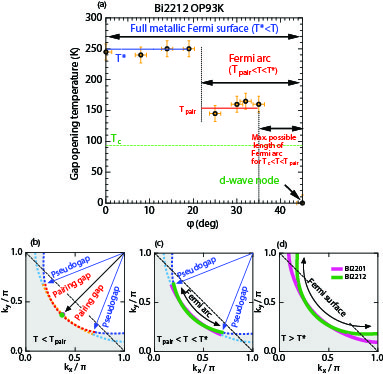}
\centering
\caption[Fig realArc] {\label{Fig realArc} New data analysis showing point nodes above $T_c$ from Ref. \cite{Kondo:FormationGaplessFermiArcs} (a) Temperature where gaps appear at different locations around Fermi surface.  $\phi=45^{\circ}$ corresponds to the node and $\phi=0^{\circ}$ corresponds to the antinode. (b)-(d) Fermiology in three temperature regimes: point nodes just above $T_c$ (b), arcs at a higher temperature (c) and a full Fermi surface above T*.  }
\end{figure}

Three analysis techniques have recently been employed towards demonstrating a single gapless point above $T_c$.  The first was Tomographic Density of States (tDOS), advanced by the Dessau group which involves taking both nodal and off-nodal cuts, integrating each cut with respect to momentum, dividing each off-nodal integrated spectrum by the nodal integrated spectrum, and fitting to a Dynes model\cite{Reber:NonQuasiparticleNature}.  Although spectra processed in this manner are less feature-ful than raw spectra, they do reveal a depression of density of states close to the node where symmetrization reports a Fermi arc, indicating a gap which symmetrization is blind to.  The second analysis technique, also relying on a more sophisticated assessment of density of states at the Fermi level, was put forth by the Kaminski group.  This analysis involves analyzing the integrated area of the momentum distribution curve (MDC) at $E_F$ as a function of temperature\cite{Kondo:FormationGaplessFermiArcs}.  This integrated area was taken to be constant when a gap was absent, and a decrease relative to this benchmark was interpreted as the opening of a gap at that temperature and momentum.  This analysis technique has the advantage of analyzing a quantity close to raw data, but it makes no claim to quantify the magnitude of gaps.  The results of this analysis were point nodes just above $T_c$ and \textit{two} characteristic temperatures above $T_c$: an intermediate temperature where gaps in the near-nodal region collapse yielding a Fermi arc and the usual pseudogap temperature, T*, where the antinodal gaps disappear.  The results of that analysis are shown in Fig. \ref{Fig realArc}.  The same intermediate temperature scale is revealed by looking at the temperature dependence of density of states at $E_F$ in symmetrized antinodal spectra\cite{Kondo:DisentanglingCooperPair}.  The third analysis procedure, put forth by T. Kondo, uses the results of the previous two sets of papers as a starting point for attempting to quantify spectral features further\cite{Kondo:PointNodesPersisting}.  This procedure \textit{assumes} that the gap closes in a BCS-like fashion at the temperature above T$_c$ established by earlier experiments, and uses the gap at each temperature as a fixed parameter while fitting $\Gamma_1$ and $\Gamma_0$ using the Norman model.  It is found that $\Gamma_0$ diverges approaching T$_c$, and $T_c$ coincides with the temperature where $\Gamma_1=\Gamma_0$, though the chosen gap closing temperature does not appear to correspond to any signatures in either parameter.

All three sets of analysis give strong support for point nodes--not Fermi arcs--just above $T_c$, though the tDOS analysis yields lower gap opening temperature.  These analyses also highlight the inherent difficulties in simultaneously and independently characterizing the gap and scattering rate in the regime where the temperature energy scale is comparable to both.  These studies make use of synchrotron, helium lamp, and laser lightsources for photoemission, which emphasizes how all three lightsources permit sufficient resolution and intensity to pursue quantitative studies of subtle spectral weight suppression.  In all the studies, they also associate the temperature above $T_c$ where the near-nodal gap closes with preformed pairs, naming that temperature $T_{pair}$.

\begin{figure}[!]
\includegraphics [type=jpg,ext=.jpg,read=.jpg,clip, width=3.5 in]{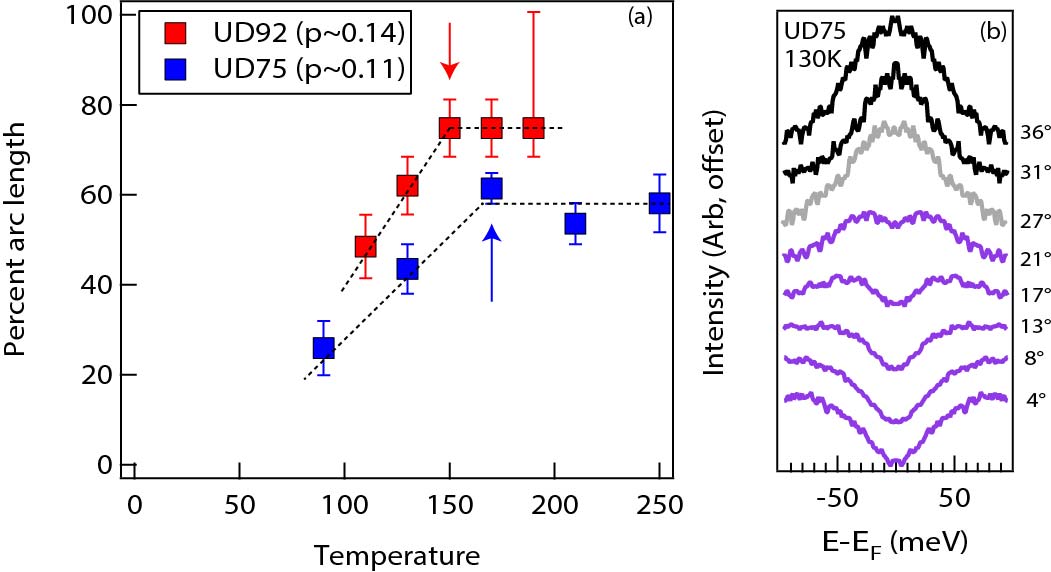}
\centering
\caption[Fig 1] {\label{Arc Length} Temperature dependence of Fermi arc for two dopings of Bi2212. (a) Arc lengths as a function of temperature from symmetrization analysis, shown in panel (b).  $0\%$ corresponds to a point node and $100\%$ corresponds to an ungapped Fermi surface achieved at $T^*$, which is higher than the measured temperatures for both dopings.  Arrows mark temperature where arc stops expanding linearly with temperature.  Note that these data dispute the conclusion of Ref. \cite{Kanigel:EvolutionPseudogap} that the normal-state temperature evolution of the Fermi arc extrapolates to point nodes at zero temperature. (b) Select symmetrized EDCs at $k_F$ for UD75 measured at 130K and offset for clarity.  The curves in black are on the arc, the one in gray is on the boundary, and the purple ones are gapped.  }
\end{figure}

The temperature scale derived by newer analysis of ARPES data described above also reveals itself through symmetrization analysis.  Fig. \ref{Arc Length} shows the Fermi arc length as a function of temperature using symmetrization, where one peak in the symmetrized EDC at $k_F$ is interpreted as being on the arc and two peaks are interpreted as being gapped.  As reported in Ref. \cite{Kanigel:EvolutionPseudogap}, which performed identical analysis, the arc length grows linearly with temperature just above $T_c$.  However, the data in Fig. \ref{Arc Length} show that for two dopings, the arc length saturates at a certain temperature.  This intermediate temperature is similar to $T_{pair}$ reported in Refs. \cite{Kondo:DisentanglingCooperPair} and \cite{Kondo:FormationGaplessFermiArcs}. At $T^*$, which is not reached in Fig. \ref{Arc Length}, a full Fermi surface is recovered, which sets the upper temperature bound of the Fermi arc plateau in Fig. \ref{Arc Length}.

The idea of preformed Cooper pairs above $T_c$ has long been part of the cuprate dialog on both theoretical and experimental grounds.  With respect to the phase diagram, one enduring controversy is the temperature above $T_c$ where local pairing becomes appreciable.  Proposals can roughly be broken into three categories: pairing at $T^*$, pairing at a temperature significantly higher than $T_c$ but smaller than $T^*$, and pairing only slightly higher than $T_c$.  The summary which follows highlights a small number of experimental results supporting each scenario.  Pairing at $T^*$ has been the favored interpretation of some spectroscopy data that show a smooth evolution of spectral gaps through $T_c$\cite{Renner:PseudogapPrecursor}, or a Fermi arc purported to extrapolate to a point node in the limit of zero temperature\cite{Kanigel:EvolutionPseudogap}.  That being said, there are examples of gap discontinuities across $T_c$\cite{Hashimoto:DirectSpectroscopicEvidence}.  Pairing at a temperature roughly intermediate between $T_c$ and $T^*$ is the favored interpretation of new analysis of near-nodal ARPES data discussed earlier and some magnetotransport experiments\cite{Rullier-Albenque:HighFieldSCFluct}, but is perhaps most prominently showcased in vortex nernst and diamagnetism measurements\cite{Wang:NernstEffect,Li:DiamagnetismCooperPairingAboveTc}, though similar measurements have been interpreted differently\cite{Cyr-Choiniere:EnahncementNernstStripe}. Finally, THz, microwave conductivity, and torque magnetometry measurements support a picture where superconducting fluctuations are significant only 10-15K above $T_c$ \cite{Corson:VanishingPhaseCoherence,Orenstein:SuperconductingFluctuations,Grbic:MicrowaveConductivityHg1201,Grbic:TemperatureRangeSuperconductingFluctuations,Bilbro:TemporalCorrelationsSuperconductivity,Yu:UniversalPrecursor}. Naturally, different experiments are sensitive to different physics, and there is no expectation for superconducting correlations to have a sharp onset.  Nevertheless, this substantial spread in purported pairing temperatures is unresolved, but one potential reconciliation of several recent results is presented below.

\section{(At least)Two temperatures above T$_c$}
Recent ARPES experiments have reported distinct spectroscopic signatures at two temperature scales above $T_c$, attributing the higher one to the pseudogap and the lower one to preformed Cooper pairs. Intriguingly, the lower one appear to coincide with the CDW onset temperature reported from scattering experiments in YBCO.  Although CDW correlations have been observed by resonant x-ray scattering (RXS) in Bi2212, their doping-dependent onsets in those experiments have not been fully reported yet\cite{daSilvaNeto:UbiquitousInterplay,Hashimoto:DirectObservationCDW}.  As discussed earlier, one should be cautious when comparing data from different experiments on different compounds, but one justification for comparing YBCO and Bi2212 data to one another comes from $T^*$ measured by ARPES and neutron scattering.  Both data are available for Bi2212 and appear to be consistent.  Additionally, neutron measurements of $T^*$ in YBCO also yield similar values to Bi-2212 with similar doping.

The first point to make about the experimental phase diagram in Fig. \ref{Fig PhaseDiagramSchematic}(b) is that the charge order appears to be distinct from the pseudogap.  The spectroscopy picture of the pseudogap, at least, how it was understood until recently, consisted of a partially-gapped Fermi surface with finite density of states inside the gap.  This shares certain features with a partially nested charge density wave with short correlation length\cite{Moore:FermiSurfaceEvolution,Chatterjee:EmergenceCoherenceCDW}.  For that reason, there was substantial excitement when a short-range CDW with wavevector $q\approx0.3$, consistent with antinodal nesting, was discovered in YBCO \cite{Ghiringhelli:LongRangeIncCharge,Chang:DirectObservationCompetition} (previously, La-based single layer cuprates had shown tendency towards charge/spin stripes at \textbf{q}$=$0.25, but this is outside the scope of this article\cite{Tranquada:EvidenceStripeCorrelations}).  However, further analysis revealed that the charge order had a distinct doping dependent onset from the pseudogap, with the former appearing as a dome peaked at $\approx 1/8$ hole doping\cite{Blanco-Canosa:RIXS_CDW,Hucker:CompetingChargeSpinSC} and the latter increasing monotonically with underdoping.  Additionally, the onset temperature, at least in compounds where it can be pinned down accurately, is consistently lower\cite{Blanco-Canosa:RIXS_CDW,Chan:CommensurateAFExcitations,Tabis:NewInsightChargeOrder,Chang:DirectObservationCompetition}.  Another important consideration for assessing the relationship between the pseudogap and charge order is that charge order has also recently been observed in electron-doped cuprates\cite{daSilvaNeto:DopingDependentCOElectronDoped} for which the partially gapped normal state is widely attributed to short-range antiferromagnetic correlations\cite{Motoyama:SpinCorrelationsElectronDoped}, different from the pseudogap on the hole-doped side.  If cuprate charge order is to be understood as being intrinsically related with the pseudogap, and vis versa, this result needs to be incorporated.

One instance where the temperature scales do match up is the `$T_{pair}$' temperatures discussed in the previous section, Nernst effect, and charge order.  The phase diagram in Fig. \ref{Fig PhaseDiagramSchematic}(b) shows various measurements of a near-nodal gap-closing temperature from ARPES experiments, Nernst onset temperatures in Bi2212, and the charge order onset from energy-integrated RXS measurements in YBCO.  The three measurements have a similar phenomenology, being peaked at $\approx12.5\%$ hole doping.  These data differ from the pseudogap temperature, $T^*$, which increases monotonically with underdoping, and from the dynamic superconducting phase fluctuation temperature, which follows the same doping dependence as the superconducting dome and tends to be much closer to $T_c$ \cite{Corson:VanishingPhaseCoherence,Orenstein:SuperconductingFluctuations,Grbic:MicrowaveConductivityHg1201,Grbic:TemperatureRangeSuperconductingFluctuations,Bilbro:TemporalCorrelationsSuperconductivity,Yu:UniversalPrecursor}.

Is the concurrence between the intermediate normal-state temperature scale often attributed to pairing and the onset of charge order purely coincidence?  Can this temperature be interpreted solely as characteristic of charge order?  Does it reflect potential connection or coexistence between the two states?  Towards the first point, it is frequently acknowledged that it is difficult or impossible to achieve a nodal state other than superconductivity without fine tuning.  Towards the second point, a number of experiments, both in cuprates and other systems where CDW exists in proximity to superconductivity ($NbSe_2$) have shown an enhancement of the Nernst signal associated with the charge order or stripe order regime\cite{Bel:NernstNbSe2,Cyr-Choiniere:EnahncementNernstStripe,Chang:NernstSeebeckCuprate}, suggesting that Nernst data in cuprates may have different interpretation than the initial one.  Towards the third point, this idea was recently explored in the context of an SU(2) order which rotates \textit{d}-wave superconductivity into \textit{d}-wave charge order\cite{Montiel:EffectiveSu2,Sachdev:BondOrder}.  Notably, the intra-unit cell symmetry of charge order was recently shown to include \textit{d}-wave character\cite{Comin:SymmetryChargeOrder}, and the onset of diamagnetism in optimally-doped YBCO happens at the same temperature as the onset of CDW ($\approx 130K$) \cite{Li:DiamagnetismCooperPairingAboveTc,Blanco-Canosa:RIXS_CDW}.  If recent ARPES signatures of an intermediate temperature scale between $T_c$ and $T^*$ can be attributed to charge order, in hole or in part, it would represent the first definitive signature of charge order in ARPES experiments on Bi2212 (consequences on the normal state band structure have been reported in Bi2201 only\cite{Hashimoto:ParticleHoleSymmetryBreaking}).

No matter the interpretation, there is now substantial experimental support for two temperature scales above T$_c$, an idea not without precedent\cite{Emery:SpinGapProximity_HTSC,Rigamonti:MainResultsNMRNQRCuprates,Timusk:PseudogapExptSurvey,Tallon:DopingDepT_star}.

\section{Conclusions}
The normal state of hole-doped cuprates may have three characteristic temperatures: the onset of superconducting phase fluctuations $\approx10-20K$ above $T_c$, the onset of charge correlations at a slightly higher temperature which is maximum near $1/8$ hole doping, and finally, the onset of the pseudogap, $T^*$.  Additionally, recent ARPES data question the appearance of Fermi arcs above $T_c$, and favor a scenario where point nodes persist to a higher temperature.  Intriguingly, this higher temperature is similar to the temperature where charge correlations onset.


\begin{acknowledgements}
The author acknowledges helpful discussions with E. H. da Silva Neto, C. Pepin, S. Chakravarty, and S. Lederer.  A portion of the data in this paper were collected in the Laboratory of Zhi-Xun Shen with support by the DOE division of Materials Sciences through SIMES.
\end{acknowledgements}

\bibliography{PG_ARPES_refs}
\end{document}